# Emergency Vehicle Preemption Strategies using Machine Learning to Optimize Traffic Operations

Somdut Roy[1], Michael Hunter[2], Abhilasha Saroj[3], and Angshuman Guin[2]
[1]AtkinsRéalis, Edison, New Jersey
[2]Georgia Institute of Technology, Atlanta, Georgia
[3]Oak Ridge National Laboratory, Oak Ridge, Tennessee

***Abstract*—Emergency-response-vehicles (ERVs), such as firetrucks, operate to save lives and mitigate property damage. Emergency-response vehicle preemption (EVP) is typically implemented to provide the right-of-way to ERVs by providing green signals to the ERVs as they approach signalized intersections along the ERV route. EVP operations are usually optimized to minimize delays to ERVs. This study strives to minimize the delay experienced by other vehicles in the network while keeping the ERV travel time near the optimum. In this paper, a machine learning–based EVP strategy, termed "MLEVP," is developed. MLEVP uses real-time sensor data, such as from vehicle detections, signal indications, and ERV location, to determine the EVP trigger time at several intersections downstream of the current location of the ERV at any point in time. MLEVP proactively clears downstream traffic/queues and minimizes ERV response time while reducing delays on conflicting traffic movements. In the case study presented, MLEVP is developed using a calibrated microscopic simulation of a signalized corridor testbed, developed in PTV-VISSIM®. In this study, the EVP problem is translated into a regression problem and then solved with machine learning techniques trained using data generated from the simulation model. Study results demonstrate the potential for the developed algorithm to produce near-optimal ERV travel time travel times while ensuring minimal impact on conflicting traffic.**

## I. Introduction

Significant strides are being made in solving real-world problems by leveraging connected vehicle (CV) and connected autonomous vehicle technology deployments [1, 2]. Such deployments result in an abundance of high-quality data that can be leveraged for real-time transportation management systems, including signal systems. One such real-time transportation application is emergency vehicle preemption (EVP), which facilitates the movement of emergency response vehicles (ERVs) with minimal delay in situations where a few minutes could dramatically change the outcome of the emergency event [3, 4]. EVP provides a GREEN indication to ERVs approaching an intersection while temporarily holding conflicting traffic movements on RED. This process not only grants the ERV right-of-way but also clears any vehicles queued in its path, ensuring the ERV can pass through the intersection without delay [5]. Traditionally, EVP systems use optical or audible cues from ERVs to clear their path. However, these methods can fail to clear the ERV path in time for uninterrupted intersection passage, especially under congested conditions. CV technology provides an opportunity to address this shortfall. Furthermore, as EVP systems target a specific set of vehicles, a successful CV-based EVP implementation is not expected to rely on a high penetration of CV technologies in the traffic ecosystem. Hence, leveraging CV, advancements in EVP are poised for near-term implementable innovation [6-8].

EVP is not a new concept and has been used in the past with varying levels of success [7, 9, 10]. However, the benefits of non-CV–based EVP has been limited in the past, especially in congested roadway conditions. This is primarily because of the line-of-sight requirement between the ERV transmitter beacon, which typically emits optical (infrared) or acoustic signals, and the preemption request receiver at the traffic signal [11, 12]. Recent technological developments, such as the integration of GPS device-based triggering mechanisms in ERV transmitters, have improved preemption reliability and reduced the system's dependence on the line of sight [13-16].

Building on previous methods, CV and connected infrastructure technology provide a seamless means to integrate live vehicle position and signal status data streams, enabling corridor-level or path-level enhanced strategies for optimal EVP performance. With an ERV's onboard CV equipment and roadside units connected to the signal controllers, traditional line-of-sight restrictions inherent to optical and acoustic EVP systems are eliminated since wireless CV messages can be transmitted even when the ERV is not in direct visual range. Although large vehicles (e.g., trucks or buses) might cause some signal attenuation, modern Dedicated Short-Range Communications and Cellular Vehicle to Everything communication standards are designed to maintain reliable transmission under such conditions, making corridor-level EVP coordination far more robust than legacy optical systems. By anticipating the arrival of the ERV, based on its position as recorded by CV messages received at other roadside units in the system, vehicles on the approach of interest can be cleared from the intersection before the ERV arrives. Although such a methodology has been proposed previously and has seen limited implementation using GPS and cellular phone-based technologies [7, 11], there is limited literature on field-based evaluations of EVP implementations involving the integration of CV technology. Limited literature also exists regarding the potential predictive methodologies required to extract information from a "connected" ecosystem enabling on-the-fly EVP decision-making. Finally, research on potential variations of EVP implementations within a CV environment is also limited.

The primary goal of this effort is to develop a machine learning (ML)–based EVP strategy, referred to as "MLEVP," that clears conflicting traffic in the path of an ERV and prioritizes minimizing ERV delay. The secondary motivation of the proposed MLEVP strategy is to manage traffic disruptions in the aftermath of an EVP call. This study is an outgrowth of previous efforts by the research team that were hindered by methods requiring real-time queue lengths, which are not easily attainable and can be highly variable (Guin et al. [17]). MLEVP seeks to use minimal real-time information and eliminates the need for rigid traffic-flow–related assumptions, such as fixed vehicle arrival rates, uniform headways, or deterministic queue discharge patterns. Instead, MLEVP relies mainly on traditional in-field detector inputs, armed with novel ML techniques, assuming connected ERVs but minimal CV penetration in the general traffic. To fulfill the secondary research motivation, experiments are performed that seek to minimize cross-street traffic disruptions resulting from an EVP event.

The study develops a simulation model for a roadway corridor in Gwinnett County, Georgia, to support all experiments. The following sections present existing literature on EVP implementation, describe the simulation model, and outline EVP experiments designed to compare the performance of a traditional Check-In Check-Out (CI-CO) EVP strategy and a previously developed Dynamic Preemption (DP) EVP strategy with the proposed MLEVP strategy. The DP ad hoc heuristic EVP strategy [16] serves as the baseline for a CV-based preemption approach. The DP heuristic assumes sufficient CV penetration to estimate real-time queue lengths. In contrast, the MLEVP strategy applies a more sophisticated approach, relying on limited CV penetration and standard intersection detectorization. The evaluation compares these strategies based on ERV travel times and general traffic performance. Additionally, a software-in-the-loop (SIL) environment demonstrates the application and performance of the developed methods.

## II. Background

Per the 2023 *Manual on Uniform Traffic Control Devices* (MUTCD), edition [18], traffic signal preemption is defined as the change in operation of a traffic signal from normal control mode to a special control mode. The primary objective is to provide uninterrupted passage to a specific class of vehicles by providing green indications along the route of the vehicle. EVP is the application of preemptive operation to ERVs (firetrucks, ambulances, police cars, etc.) with an aim to give right-of-way to the ERV, minimizing the delay in reaching the incident location [19].

### *A. Strategies for Effective EVP Systems*

Depending on the vehicle type that requires right-of-way, the logic of a preemption algorithm can vary. Many factors govern the preemption call algorithm, including location, vehicle type, time of day, and traffic demand levels. Existing literature on EVP identifies common practices, with the primary objective being to extend the green phase to reduce ERV travel time and ensure a safe and clear pathway for the vehicle [20].

The preemption process involves two distinct transition phases: entering the preemption state and exiting preemption to restore normal signal operations. The *Traffic Signal Timing Manual* [21] states that for both transitions the yellow and all-red intervals shall not be shortened or omitted. Additional transition guidance addresses pedestrian timing constraints, returning to a red indication, allowable indication transitions, and handling multiple simultaneous preemption requests. In the current study, several of these transitions are explored under different experimental setups to examine their functionality and impact on both the ERV and conflicting traffic. To better understand the operational characteristics and limitations of these methods, two distinct EVP strategies are evaluated: CI-CO and DP.

**CI-CO**: In a traditional CI-CO configuration, a check-in detector is situated at a predetermined distance from the intersection. When the ERV reaches this detector, a preemption call is initiated, which remains active until the vehicle passes a check-out detector (or times out). The check-out detector is typically located immediately downstream of the intersection. A limitation of CI-CO arises when an ERV encounters the rear of a queue that extends beyond the check-in detector. In such cases, the preemption call is not activated until the queue has been sufficiently cleared for the ERV to reach the check-in detector. Furthermore, this method does not account for the number of vehicles queued between the ERV and the intersection, potentially leading to inefficient operation. Even more recent applications of CV in EVP tend to follow the CI-CO strategy, creating a set distance in a MAP file for the activation of preemption rather than dynamic call placement based on traffic conditions.

**DP**: The DP strategy [22] was developed by the authors of this effort. DP is a forward-looking EVP strategy that determines whether to trigger preemption before an ERV reaches an intersection using real-time data streams rather than relying on fixed-distance triggers or optical systems. This method is evaluated within a digital twin testbed that replicates a corridor of signalized intersections on an urban arterial in Georgia, providing a realistic simulation environment. DP aims to strike a balance of maximizing reductions in ERV travel time while limiting the negative impacts on cross-street traffic. Results indicate that the DP approach achieved a substantial 20% reduction in ERV travel times compared with traditional CI-CO EVP practices. By leveraging a digital twin environment and live traffic data, DP demonstrates a promising direction for more efficient, corridor-level preemption strategies in urban settings. However, DP has significant data requirements in that queue estimates are needed.

### *B. Use of AI/ML-Based EVP*

In addition to the novel ML and AI concepts being developed to support self-driving vehicle implementations [23-27], the application of ML to understand aspects of travel behavior and decisions in transportation has an extensive history [28]. The following focuses primarily on studies related to EVP. For example, a detection study by Islam and Abdel-Aty (2022) proposes a support vector machine (SVM)-based strategy to detect an ERV in real time with the aid of audio data, achieving high accuracy [29]. A recent study performed by Pan et al. (2022) addresses the arrival of ERVs in congestion scenarios using a dynamic method that applies Kullback-Leibler divergence to measure traffic density and prepare a corridor for the EVP call [30]. Savithramma et al. (2022) investigate a problem closely aligned to the current effort by employing convoluted neural networks to predict the required green time at a given signalized interchange [31]. A study by Shaaban et al. (2019) seeks the optimal initiation point for the start of the green phase, accounting for delays to conflicting traffic [19]. It also considers the optimal path suggestion for the ERV. The study assumed significant vehicle-to-vehicle and vehicle-to-infrastructure technology penetration. The study suggests extensions to the effort such as (a) using real-time feedback from the vehicle fleet (in the form of detector pings) and (b) leveraging ML techniques to address more complex, real-world incident scenarios.

These efforts provide essential considerations that help form the basis of the current effort. In addition, several other EVP studies that leverage novel ML techniques are tangential to the current effort: for example, (1) designing EVP under conditions of uncertainty in road incidents [32] and (2) surveys addressing EVP challenges based on routing and scheduling [33]. These studies provide additional motivation for the current experimental design.

## III. MOTIVATION

The existing literature emphasizes the importance of studying the impact of EVP by accounting for travel times on the route of the ERV as well as the intersecting routes that are interrupted. Previous studies have evaluated algorithms designed to optimize EVP performance and minimize traffic disruption. However, these studies are typically limited to preemption at a single intersection or a few noncontiguous intersections within a small network.

The effort presented here focuses on addressing the EVP challenge on a medium-sized network consisting of 25 signalized intersections, 14 of which are located along the EVP corridor. The study implements preemption on eight contiguous intersections on the main line, focusing on the challenges related to main line signal coordination disruption and side-street traffic impacts. The case study investigates preemption strategies during the PM peak hour (5 PM to 6 PM), which is typically the most congested period for this corridor. The generalized MLEVP solution leverages traffic parameters that can be reasonably collected in real time. The MLEVP thus aims to make EVP decisions based on real-time information.

### *A. Overview of Proposed Methodology*

To counter the limitations of the DP EVP approach mentioned in the previous section, a supervised ML-based approach, MLEVP, was developed. The proposed MLEVP strategy is designed to be implemented on a smart corridor equipped with sensors to obtain (a) real-time vehicle detection upstream of the stop-bar (ideally, just downstream of the previous intersection) and the ERV GPS location and (b)signal state data including Signal Phasing and Timing (SPaT). In addition, implementing the MLEVP strategy requires the capability to modify the signal preemption parameters in the controller in real time.

The goal of MLEVP is to identify the point in time beyond which providing an earlier preemption does not significantly improve emergency response time. An MLEVP model was developed for each intersection along the corridor. The MLEVP model for each intersection considers **multiple ML algorithms: (1) support vector regressor (SVR), (2) neural network regressor, and (3) random forest regressor**. Implementing all of the intersection-level MLEVP models collectively constitutes the MLEVP strategy.

The development process for MLEVP strategy involved three primary steps:

1. **Simulation-based training data generation:** First, using simulation, optimal preemption times are determined for different scenarios of ERV entry times into the network. From these runs, the following features and labels are extracted to be used as training data. Features and labels are determined for each second of simulation.
   - **X (Features):** Traffic state variables such as signal status, preemption status, detection status, and ERV distance from the signal over previous time intervals.
   - **Y (Label):** EVP trigger, which will be seen to be a soft label ranging over [0,1] rather than strictly 0 (no preempt call) or 1 (preempt call).
2. **Intersection-level model development:** Using the generated training data, each model is trained, validated, and tested to determine the optimal time to trigger an EVP call. Integrating these models across all intersections forms the MLEVP strategy for the study area.
3. **Performance evaluation of the MLEVP strategy:** comparing its effectiveness against traditional CI-CO and DP strategies.

The next sections discuss each of the preceding in more detail.

*B. Optimal Solution-Based Training Data Generation Through Simulation*

1) **Simulation Model Development**

A microscopic simulation model was developed in Norcross, Georgia, 6.2 miles along the Peachtree Industrial Boulevard (PIB) corridor, from Holcomb Bridge Road at the southwest end to Pleasant Hill Road on the northeast end. The model includes 25 intersections along and adjacent to PIB. The layout of the network in PTV VISSIM® [34] and in satellite view is shown, respectively, in Fig. 1. For consistency in naming, the PIB directions of travel are considered north and south throughout the length of the corridor for the remainder of this paper. The cross-street approaches are defined as eastbound and westbound.

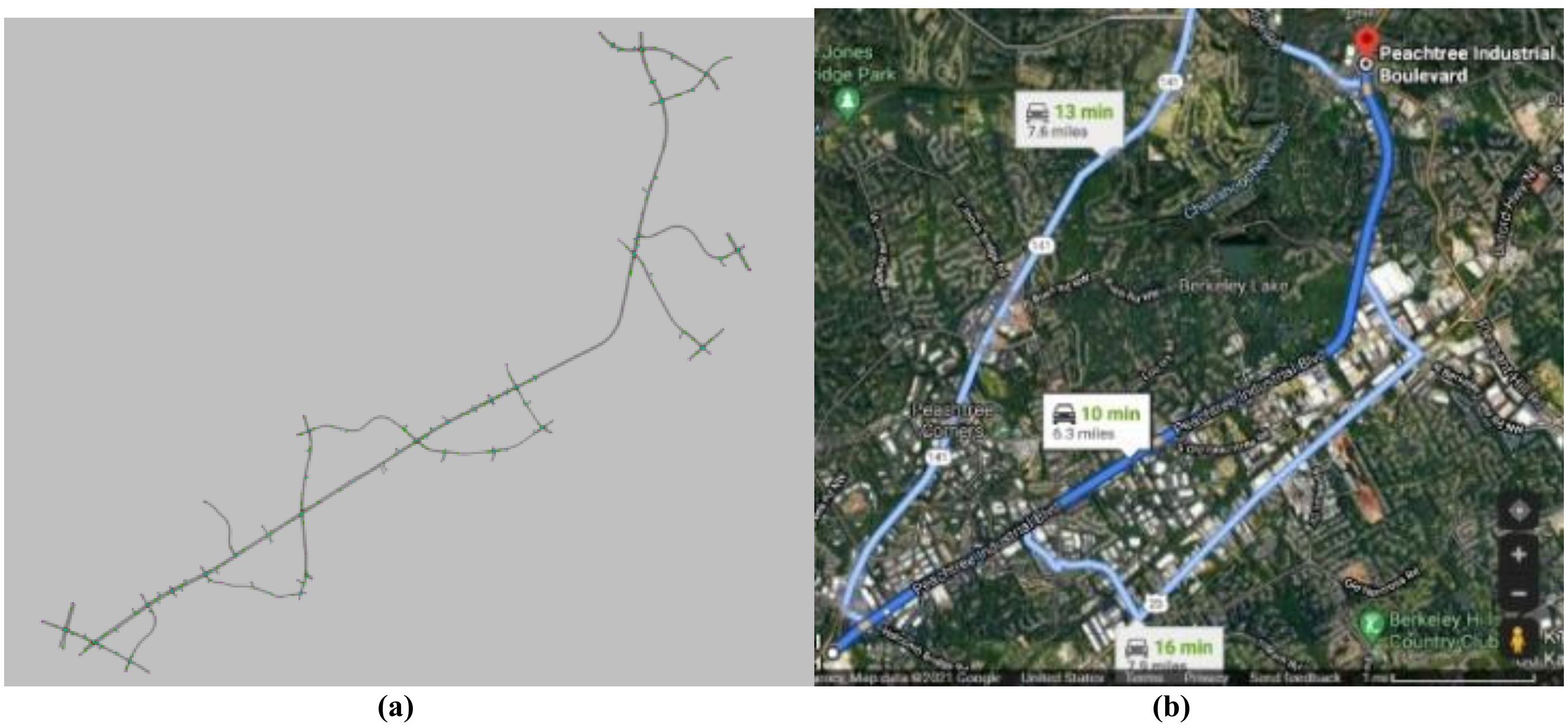

**(a)** **(b)**

**Fig. 1.** Case study network of PIB: (a) PTV-VISSIM® simulation model and (b) satellite view by Google Maps™ [35].

2) **Optimal Preemption Time Determination**

The training data comprises traffic state variables (features) that represent the external environment, along with the corresponding optimal preemption time for the ERV in that environment (target variable). The developed simulation model is employed to generate these features and target values across a range of scenarios.

To determine the optimal preemption time, the simulation is used to assess the performance of both the ERV and background traffic under varying preemption call placements. The optimal preemption time is defined as the closest position to the intersection at which the lowest ERV travel time is achieved and beyond which triggering the EVP earlier yields no significant improvement in response time. An iterative binary search algorithm is employed to determine the optimal placement.

The iterative binary search begins with two boundary conditions in the search for the optimal EVP call placement: (1) place the EVP call immediately upon ERV entry into the network, representing the earliest possible placement, and (b) place the EVP call once the ERV has crossed the upstream intersection and is within 1,000 ft of the downstream stop-bar, a common placement of traditional CI-CO static systems. The binary search method iteratively adjusts the preemption timing based on observed trends in ERV response times, seeking the point where further earlier preemption does not reduce ERV response time. This point is identified as optimal because initiating preemption earlier would not enhance ERV travel time and would unnecessarily increase delays for traffic in conflicting movements across the ERV path. Further, by setting the 1,000 ft upstream limit for a preempt call this implementation seeks to determine when an earlier preempt call than that of the CI-CO system could prove beneficial while guaranteeing the ERV the preempt service of the traditional CI-CO system at a minimum. If an agency desires, this limit could be set lower, potentially to the stop-bar implying no preemption call; however, this must be weighed with the potential for a poor prediction, failing to provide a call when one is needed or providing a call later than needed.

To generate the training data, signal state constraints, such as ensuring minimum green times on active phases, were incorporated to accurately reflect real-world signal operations. A total of 32 scenarios are created by introducing the ERV into the network at successive 5 s increments across the 160 s global signal cycle. To incorporate stochastic variability, each scenario is replicated five times using different random seeds. In total, this yields $32 \times 5 = 160$ simulation runs per experiment, covering all combinations of ERV introduction times and random seeds.

3) **Training Data Extraction**

The training dataset or model input matrix comprises a comprehensive array of features that can be extracted in real-time from a network, serving as critical inputs for the ML algorithm. This algorithm leverages these features to optimize the timing of EVP calls at signalized intersections. The system processes states and executes models at a frequency of once per second. The feature set includes the following:

- Location of the ERV with respect to the signal head (feet).
- Signal state of the signal under study and adjacent signals (red, green, yellow).
- Preemption state of adjacent signals (active or not, i.e., 1 or 0).
- Advanced detectors count data.
- Elapsed time since ERV enters the corridor.

The advanced detection—field installed in the range of 200–500 ft upstream of the stop-bar in the main-line PIB—provides a surrogate measure for the traffic volume on the ERV path during the time of the emergency response call. The elapsed time since entering the corridor provides the model information on the progression of the ERV. In this effort, elapsed time is from when the ERV enters the simulated network. The equation to be solved is

$$\hat{}P = f\left( \{ S_i \mid i \in [1, N], t \in [0, T]\}, \{P_j(t) \mid i \in [1, M], t \in [0, T]\}, \{ D_{i(t)} \mid i \in [1, N], t \in [0, T]\}, d, \tau \right) ,$$

where

| | |
|---|---|
| $\hat{}P$ | Preempt status for the current signal. |
| $T$ | Seconds worth of historical data used to train our model (i.e., 160 s in this case, which was the global signal cycle length). |
| $N$ | The number of intersections being considered for preemption. |
| $M$ | The number of intersections being considered for preemption downstream of current ERV intersection. |
| $S_i(t)$ | Signal state for signal *i* at time *t* ,<br>where 0 < t < T. Thus, t = T is the current signal state, t = 0 is the time of entry of ERV. |
| $P_j(t)$ | EVP state for downstream signal *j* at time *t.* |
| $D_i(t)$ | Detector state for signal *i* at time *t.* |
| d | Distance to the intersection or ERV distance to the signal. |
| $\tau$ | Time elapsed since the ERV entered the corridor to be considered for EVP. |

$\hat{}P$ is a value that would either be 0 [preemption OFF] or 1 [preemption ON]. Hence, in the initial ML development, the EVP state of a training data instance (i.e., the EVP state is whether the preemption call has been placed) was set to be "zero" before the EVP is triggered and "one" when EVP is triggered. Once the preemption is called no additional data is collected. However, this approach led to a classification problem because of a significant imbalance in the number of 0s and 1s, as after many seconds of no preemption [i.e., preempt status = 0] through which the model needs to be trained, a "1" is seen when EVP is called. Hence, a ML model that never predicts an EVP will have 100% 0s, resulting in a high accuracy, but will not provide any EVP. This is called the **accuracy paradox** in imbalanced classification. Also, a 0 or 1 system does not provide any information to the system about the need for EVP increasing with time as the ERV approaches a given intersection.

To address these issues, a soft-labeling approach [36] is adopted. This approach labels the training data from 0 when the ERV enters the network to up to 1 (see following threshold discussion), when the EVP needs to be triggered at an intersection for optimal benefit. Figure 2 displays several potential soft-labeling curves, each monotonically increasing (indicating progressively increasing need for EVP with time) to reach a predetermined *cutoff* at which the EVP is to be triggered at the intersection in question. If perfect knowledge is available and there was no variability in the EVP decisions, the *cutoff* should be 1 as this is the last moment for implementation of EVP without impacting the ERV. However, a threshold value of less than 1 allows some robustness in the system, ensuring a necessary EVP call will be made even in the event of some uncertainty. Thus, rather than training the ML to directly predict the preemption time, the ML is trained to predict the soft-label values. Once this value reaches the prescribed threshold, preempt will be called.

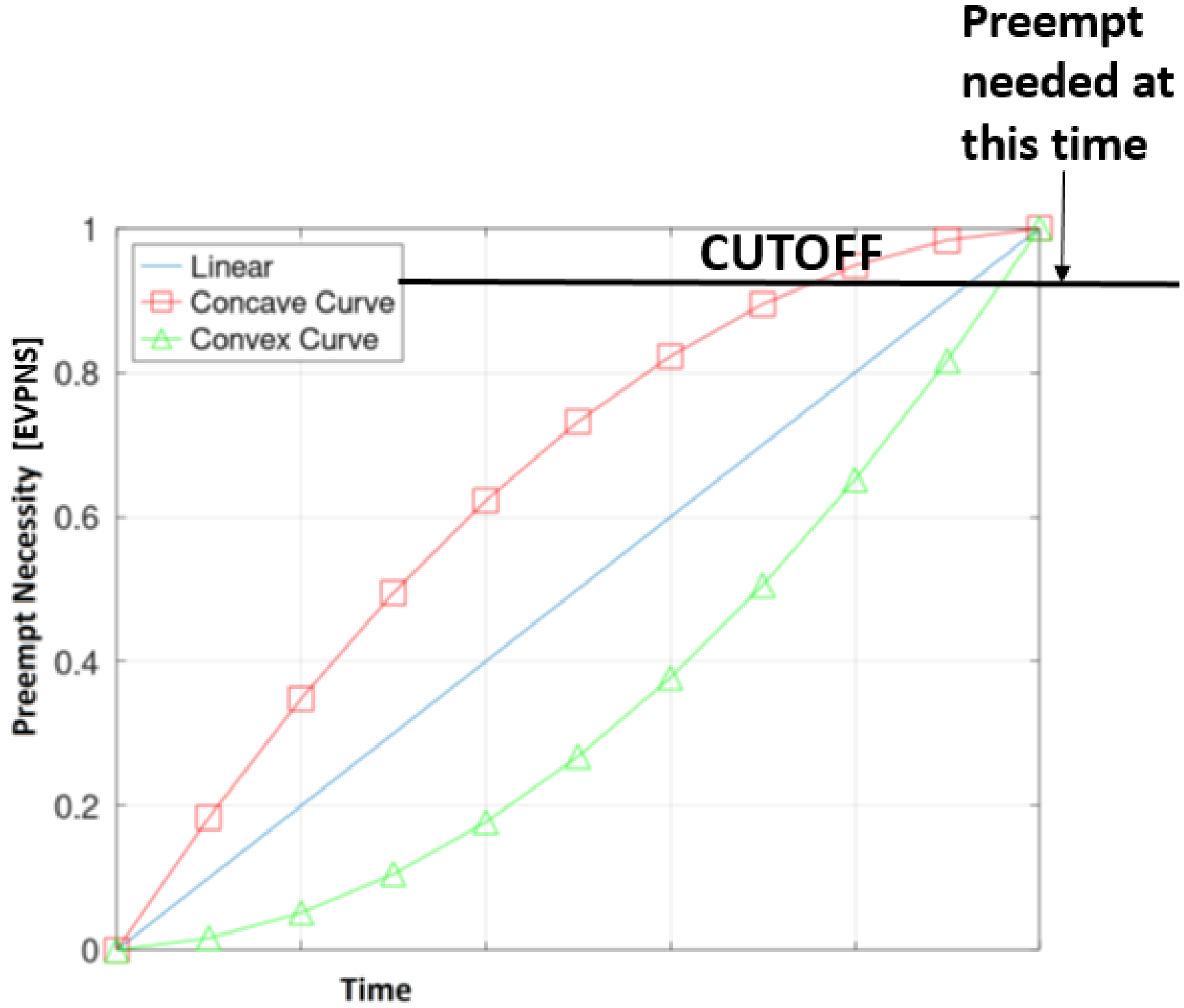

**Fig. 2.** Demonstration of soft-labeling decision-making where EVP call is needed showing combination of linear and nonlinear curves.

a) Training Data Labeling Using a Soft-Labeling Solution with Respect to EVP

The soft-labeled value is intuitively a measure of the need for EVP at a given point in time, which can be referred to as an **"EVP Necessity Score" or EVPNS**. Because there is no literature on the nature of the EVPNS curve with time, several forms of monotonically increasing curves were chosen for testing, including linear and convex or concave forms of exponential, sigmoid, and negative-exponential. The EVPNS curve is used to fit each training data set (i.e., simulation output), with values monotonically increasing from 0 to 1 according to the given form. When EVP is needed before the minimum setback (1,000 ft) described earlier, the EVP prediction model identifies when the call is to be made using a *cutoff* threshold. Note that not all ERVs require an EVP call before the selected minimum value. For instance, no call would be needed when the ERV enters the upstream approach when the vehicle can cross the intersection on a GREEN indication without preemption. When placing a call further upstream than the selected minimum does not provide a benefit, the soft-label values will range from 0 to *NoPrThres*, a value lower than the EVP call *cutoff* threshold. Hence, the following three parameters are set as part of the soft labeling:

1) ***Cutoff*:** The value (<1) that determines when EVP will be called. If the EVPNS at time *t* exceeds the *cutoff*, an EVP call is made. Selecting the cutoff is discussed in a subsequent section.
2) **No Preempt Threshold (*NoPrThres*)**: In cases where no preempt is needed before the selected minimum call distance, the EVPNS curve will go from 0 to this value (which is less than *cutoff*). This allows the model to learn cases where the EVPNS does not reach the EVP *cutoff*, thus not triggering an early EVP call. In cases that need EVP, the *NoPrThres* parameter will have a value of *1* by default. The time at which the soft label reaches *NoPrThres* is when a preempt is placed because of the ERV reaching the minimum distance (1,000 ft upstream of the intersection in this case), rather than placing the call based on ERV travel time.
3) **The form of the EVPNS curve:** This is the form of the monotonically increasing curve with respect to time, which might be linear or other traditional nonlinear curves, from 0 to 1.

Figure 2 and Fig. 3 demonstrate the functionalities of the three soft-labeling parameters. Each second of the simulation training represents a monotonically increasing value of EVPNS. In Fig. 2 it is observed that the value "1" represents the optimal EVP time, with a cutoff for calling EVP set slightly lower. An EVPNS curve could be linear, convex, or concave, but it would monotonically reach and exceed the predetermined threshold value. The moment the predicted EVPNS value reaches the cutoff, an EVP call would be made at the intersection programmed with the EVPNS curve. Fig. 3 shows a situation in which no EVP call is needed, with *NoPrThres* being reached before the cutoff.

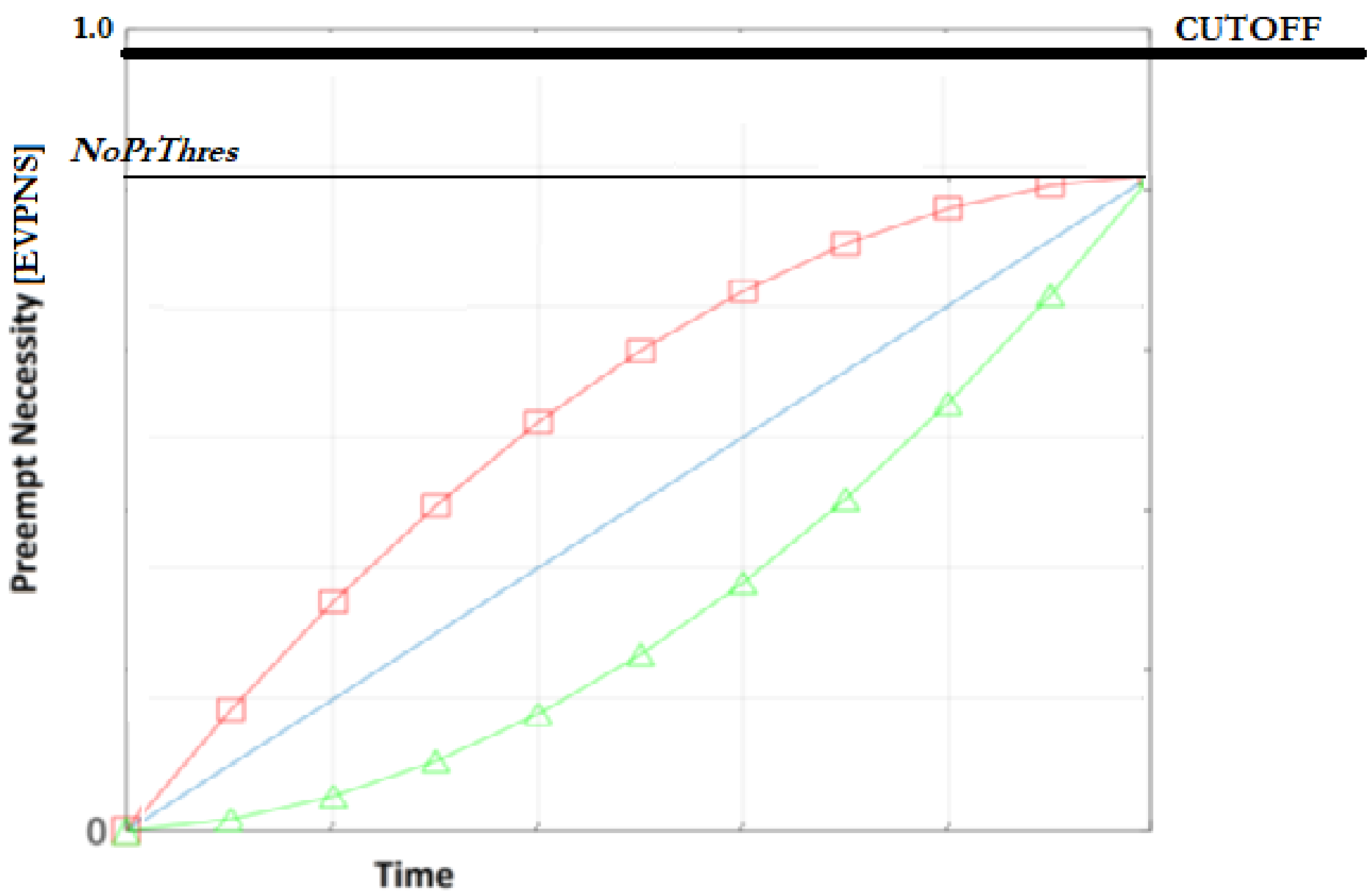

**Fig. 3.** Demonstration of use of soft-labeling parameters where EVP call is not needed before the minimum distance and *NoPrThres* < *cutoff*.

b) Input Matrix Formulation for Model Training

The previously discussed data elements are used together to create the training data input matrix for the ML model. Fig. **4** (a) shows how the features are arranged to form the input vectors used to train the ML model. The model is trained to decide whether an EVP call needs to be made, by predicting the EVPNS, for each given traffic signal controller. When the predicted EVPNS exceeds the cutoff, a preemption call is placed.

The input matrix for an intersection is filled from the time when the ERV enters the network to the time an EVP call is made at the intersection. If in a scenario an earlier EVP call is not needed, the input matrix uses the data from the time of the ERV entry until the ERV reaches a minimum distance of 1,000 ft to the intersection where a preempt call is placed by default in this implementation, as discussed previously. Using Fig. 4 (b) as reference, the preempt output will be 0 from the time the ERV enters the network and will become 1 at the time stamp when the EVP call is placed. Hence, the aim of the model is to predict the time at which the necessity for preemption exceeds the cutoff and the EVP call is placed. Once the EVP call is made, the call stays active at the intersection until the ERV crosses the stop-bar. Hence, training data stops when the call is placed.

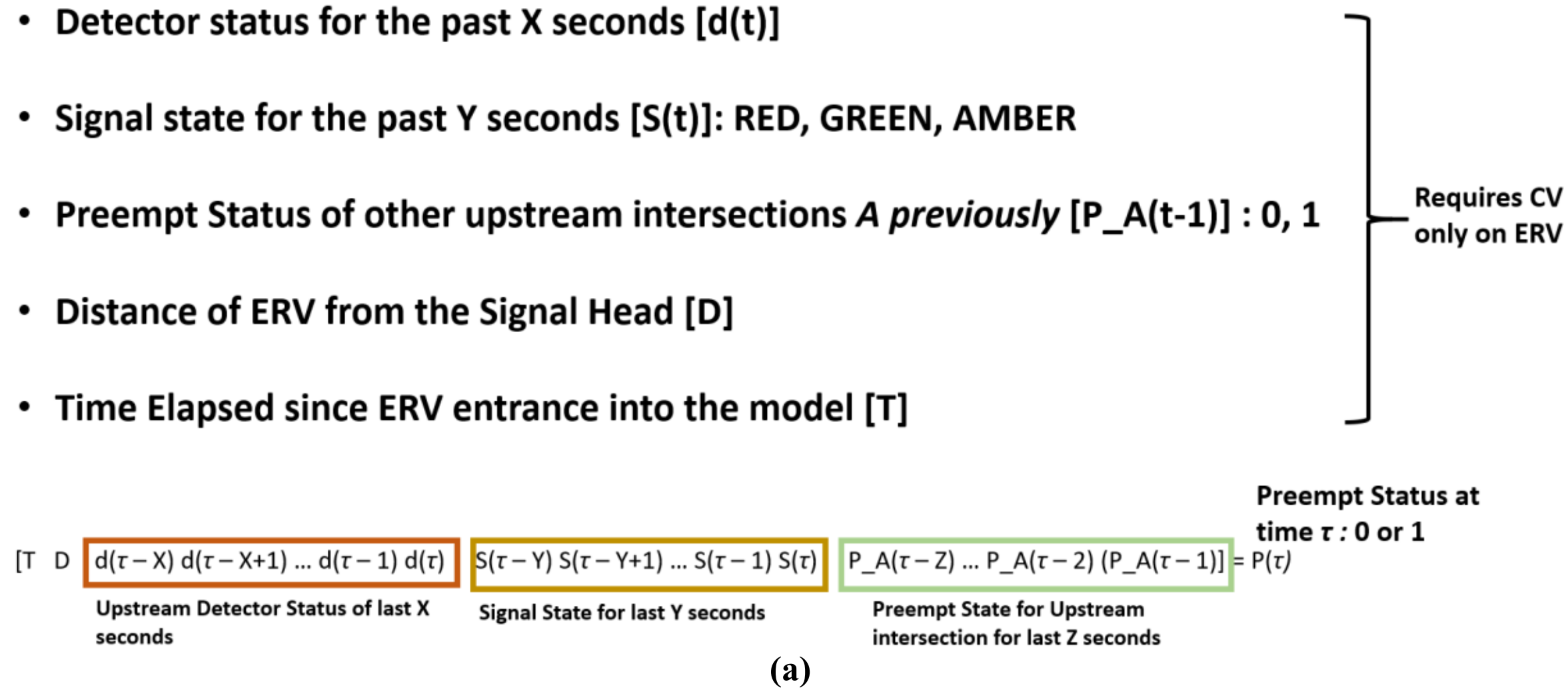

(a)

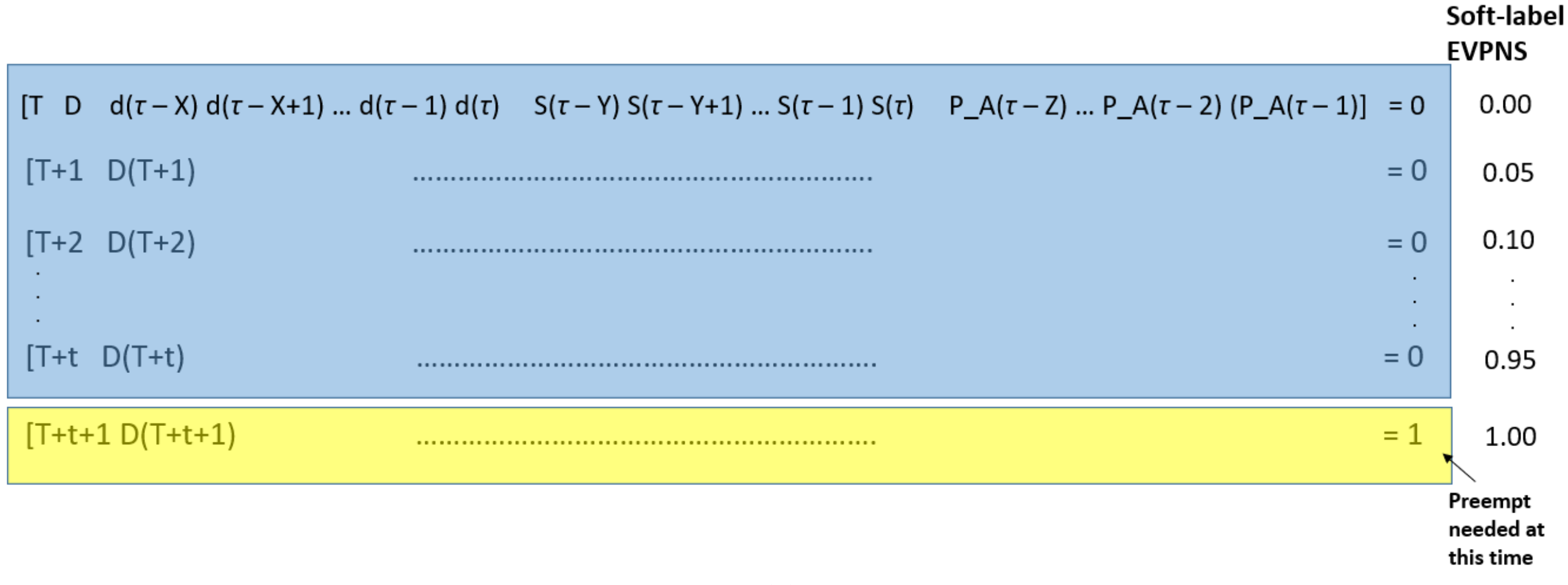

**(b)**

**Fig. 4**. Schematic representations of development of input matrix: (a) input vector of a PIB intersection and (b) input matrix and corresponding EVP output for a PIB intersection.

c) Splitting Dataset into Training, Testing, and Validation

The initial step in training a supervised learning model involves dividing the dataset into training, validation, and test subsets. Although there are various standard approaches for splitting the data [37], it is important to consider specific factors relevant to the problem before finalizing the split. In this study, 160 different scenarios were generated for each intersection, representing a range of cases from those requiring early preemption to those in which minimal preemption was needed. These variations depended on traffic variability and the ERV arrival time relative to the traffic signal cycle. To ensure that each subset adequately represented the diversity of scenarios, the data was intentionally split into a 60-20-20 ratio. Specifically, 96 scenarios were allocated for training, 32 for validation, and the remaining 32 for testing.

*C. The "Ideal" Relaxed-Optimal Solution*

When developing EVP strategies, we found that the strict optimal solution can create practical challenges. For example, a very small improvement in ERV travel time (e.g., on the order of 1–2 s) might require the preemption to last much longer, which causes significant unnecessary delays for conflicting traffic and potentially introduces its own safety challenges with extensive unexpected cross street queueing. To enable some minimal flexibility, we rely on a relaxed-optimal solution. This method accepts any preemption timing that results in ERV intersection-level travel time within a small margin, such as 2 s of the optimal. We implement this using an iterative binary search. By focusing on the relaxed-optimal solution, we maintain nearly optimal ERV travel times while reducing unnecessary delays for other vehicles, making the strategy more suitable for real-world operations. For convenience in the remainder of this paper, we refer to this relaxed-optimal solution as the "ideal" solution.

*D. Intersection-Level MLEVP Model Development*

1) **Model Training**

Multiple combinations of *NoPrThres*, EVPNS curve, and *cutoff* were used to create a hybrid model. A supervised learning approach is adopted for predicting EVPNS at a given time. Multiple ML algorithms were applied and tested: (1) SVR, (2) neural network regressor, and (3) random forest regressor. For each of the model training processes, k-fold cross validation was performed to choose the best performing hyper-parameters and mean-squared error was used as the loss function to reduce the propensity to overfit and biases [38]. In some cases there were very few data points for training, for example, where EVP is triggered soon after the ERV entry into the network. Hence, cross-validation was preferred over holdout to allow the model to train for various train-test splits, allowing for more well-informed decisions on unseen datasets [39]. For all ML procedures, Python's Scikit-learn Package was used.

2) **Model Testing—Choose Intersection-Level Prediction Model**

For each intersection, multiple trained models were developed with a combination of soft-labeling curve forms (linear, convex exponential, concave exponential, quadratic), ML algorithms, and *Cutoff* and *NoPrThres* thresholds. These combination models were trained on the training data previously described and were tested on the validation dataset. Choosing a model for a given intersection requires accounting for the parameters that are a priority when placing an EVP call. Primarily, two items factor into the final model selection:

(1) Duration for which EVP call is made but was NOT needed (i.e., false positive).
(2) Duration for which EVP call was needed, but the model did NOT activate EVP (i.e., false negative).

The former dictates how much extra EVP is given by the model in question, which in turn will cause excess delay on the conflicting movements, the reduction of which is a secondary study objective. The latter indicates a missed or delayed EVP call, which can cause delay in ERV response, the reduction of which is the primary objective of the study. Hence, to be consistent with the primary objective of the analysis, which is prioritizing sufficient preemption time, false negatives are weighted ten times more than the false positives. This ratio is selected to reflect the importance of providing minimal ERV travel time over excess conflicting movement delay. This ratio could be varied based on model developer or agency priorities. Lastly, true positives (i.e., an EVP call was needed and was made correctly) are used to normalize the scores. Hence, the selected model will have the lowest weighted factor defined as follows:

$$\boldsymbol{EVP\ model\ score} = \frac{\mathbf{0.1 * \mathit{false\ positives} + \mathit{false\ negatives}}}{\mathbf{\mathit{true\ positives}}} \ ,$$

given that

| | |
|---|---|
| $\boldsymbol{false\ positives}$ | Instances where preemption was not needed but the model predicted preemption |
| $\boldsymbol{false\ negatives}$ | Instances where preemption was needed but the model failed to predict preemption |
| $\boldsymbol{true\ positives}$ | Instances where EVP was needed and correctly predicted |

In cases of ties, lower run time and the lower delay to the conflicting traffic were used for the final selection.

*E. Combining the Intersection-Level MLEVPs for the Corridor MLEVP Strategy*

In the evaluation of intersection-level models for EVP, the neural net regressor consistently emerged as the most effective model across all intersections. In contrast, SVR did not deliver promising results for any intersection. Simple logistic regression demonstrated performance comparable with the neural net regressor only at the North Berkeley Lake Road intersection, with suboptimal results at other intersections.

SVR's limitations are influenced by factors such as (1) large datasets, (2) datasets where the number of features exceeds the number of training samples, and (3) noisy datasets [40]. Although the simulation-based training dataset exhibits lower noise levels, items (1) and particularly (2) contribute to SVR's drawbacks. The training matrix incorporates numerous features, capturing data from the last 160 s as individual features at any given simulation second. Consequently, the available data is insufficient to adequately train SVR. If training data were generated with additional random replications, resulting in a larger number of training samples, reconsideration of SVR might be warranted. Additionally, SVR's training time was significantly longer compared with both the logistic regression and neural network algorithms. Lastly, the spacing between all intersections on the ERV route, excluding one, were relatively close (less than or equal to 3,000 ft). The interaction resulting from the close spacing likely hampered the effectiveness of the linear model.

At the system level, the EVP problem is addressed for a designated ERV route by integrating solutions from individual intersection models. An EVP model is developed for each intersection, and once the ERV specifies its intended route, the intersections along that path concurrently activate their respective intersection-level EVP recommender models to determine the optimal EVP trigger times. For an ERV route involving n intersections, the models function in tandem according to the following equation:

$$T_{\text{end}} = \min(T \mid \exists i \in [1, n], M_i(T) > C_i) \ ,$$

where

$T_{end}$: Time when the simulation stops predicting.

$M_i(T)$: Output value of model $M_i$ at time T.

$C_i$: Cutoff threshold for model $M_i$.

n: Total number of models (one per intersection).

T: Current simulation time.

TABLE I summarizes the best intersection level ML models. Neural net regressor is the preferred EVP recommender for all intersections. The sample code and instructions for executing the training and validation process are available at the GitHub repository here [40].

TABLE I
SUMMARY OF THE FINAL INTERSECTION LEVEL ML MODELS FOR THE SEVEN CONTIGUOUS INTERSECTIONS INVOLVED IN THIS STUDY

| **Intersection** | ***NoPrThres*** | **Soft-labeling function** | ***Cutoff*** |
|---|---|---|---|
| South Old Peachtree Road | 0.95 | Positive exponential | 0.94 |
| Highwoods Center | 0.95 | Linear | 0.95 |

| Pickneyville Park | 0.925 | Negative exponential | 0.925 |
|---|---|---|---|
| South Berkeley Lake Road | 0.93 | Linear | 0.93 |
| North Berkeley Lake Road | 0.9 | Negative exponential | 0.9 |
| Summit Ridge Road | 0.98 | Sigmoid | 0.9 |
| Howell Ferry Road | 0.9 | Positive exponential | 0.9 |

## IV. MLEVP Strategy Performance Comparison with Traditional EVP Strategies

Each intersection has its own MLEVP model that has been built and tested individually at the intersection level. The following analysis examines how these models perform when implemented collectively at the network level. The results are derived from applying the MLEVP strategy to the 20% of scenarios that were set aside as the test set. This evaluation focuses on the overall model performance across a range of test scenarios, including cases with variations in traffic volume and ERV arrival time, as well as assessments under the SIL architecture. Here, the SIL architecture refers to a simulation environment in which actual traffic signal controller software, identical to that used in the field, is integrated with the traffic simulation model, allowing for a more realistic and practice-ready evaluation of the preemption strategies. Rather than conducting a sensitivity analysis, which would specifically involve systematically altering individual factors to observe their impact on model performance, this section presents a summary of the MLEVP model's effectiveness across the diverse conditions represented in the test set. The outcomes provide insight into the model's robustness and generalizability under realistic operational scenarios.

### *A. Performance Summary of the Model*

The following are the six experimental setups that were used in the performance comparison:

1. No EVP: No EVP is applied; signals operate under normal conditions.
2. CI-CO: CI-CO preemption triggers when the ERV enters and exits fixed detector locations near intersections.
3. DP: Uses real-time data to trigger preemption based on estimated queue clearance needs, as described by Roy et al. [22].
4. Optimal solution: Strictly minimizes ERV travel time using all available data, often resulting in longer preemption durations and increased delays for conflicting traffic.
5. Ideal solution: Refers to the relaxed-optimal solution, which accepts preemption timings that achieve ERV travel time within a small margin (2 s) of the minimum.
6. MLEVP: Predicts preemption timing using real-time traffic and signal data for practical, near-optimal ERV performance.

Fig. 5 (a) compares ERV average speeds across the six experimental setups. As expected, No EVP (orange) shows the lowest average speed at 31.4 mph, reflecting standard signal operations without preemption. CI-CO (green) uses fixed detectors for preemption and improves the average speed to 34.4 mph. The DP solution (magenta) applies dynamic preemption based on real-time data and achieves 39.3 mph average ERV speed. The optimal solution (teal) strictly minimizes ERV travel time using all available data, reaching an average of 43.4 mph, but often with long preemption durations. The ideal solution (grey) refers to the relaxed-optimal approach, balancing ERV travel time with practical constraints and resulting in an average speed of 41.6 mph. MLEVP (yellow) uses ML to predict preemption timing and achieves an average of 42.5 mph that falls between the optimal and ideal solutions. The violin plots illustrate the distribution of speeds for each strategy, with dashed lines marking the 25th and 75th percentiles and a red box indicating the mean. Fig. 5 (b) presents similar violin plots for ERV travel times and provides a complementary view of how each strategy performs in minimizing travel time. Clearly, that the optimal, ideal, and MLEVP strategies provide similar travel times, dominating the other alternatives.

Overall, the MLEVP model demonstrates superior performance in optimizing ERV passage through the corridor, aligning closely with relaxed optimal and optimal solutions, and outperforming traditional methods like DP in terms of both travel time and ERV speed outcomes.

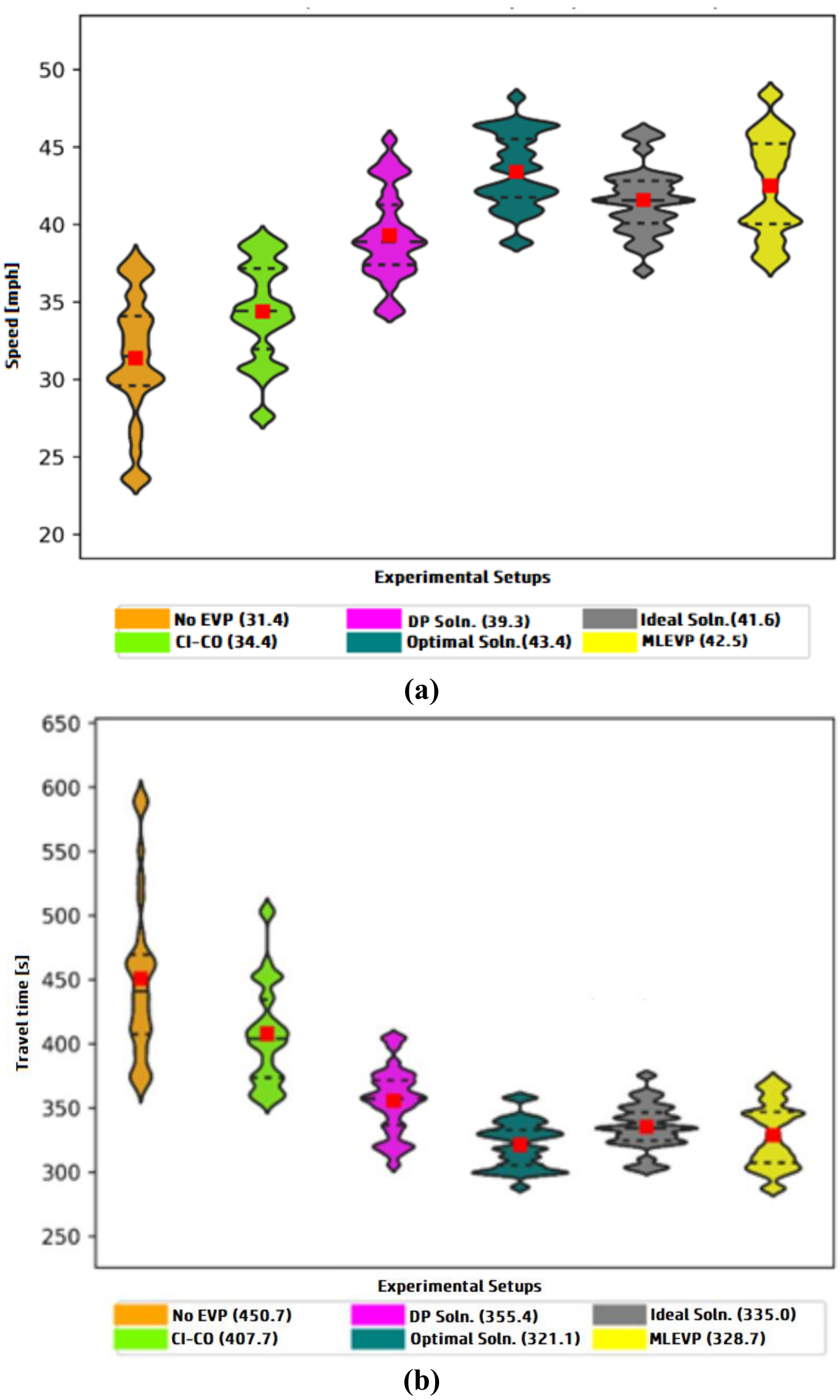

**Fig. 5.** Comparative results in violin-plots: ML prediction model vs. other EVP call experiments on the test data: main-line PIB route: (a) ERV speed (miles per hour) and (b) ERV travel time (seconds).

Fig. 6 presents the aggregate preempt duration across the seven interchanges along PIB for different experimental strategies, demonstrating that the ML-based EVP model achieves total preemption durations closely matching the ideal solution, with no statistically significant difference between the two. This indicates that the ML approach can efficiently reduce unproductive preemption time while maintaining ERV priority. Complementarily, Fig. 7 illustrates the variation in non-ERV vehicles' travel times with respect to ERV trajectory and preemption events. Figure 7 (a) shows travel times for non-ERVs along the main-line PIB, while Fig. 7 (b) shows aggregates of travel times for all side-street vehicles following an EVP event. The ML-based EVP model closely tracks the ideal solution in both mean and median travel times for non-ERVs, particularly on side streets, and outperforms the DP algorithm that produces significantly higher side-street delays. Together, these results highlight the advantage of the MLEVP approach in balancing ERV efficiency with the mitigation of negative impacts on general traffic, especially on side streets.

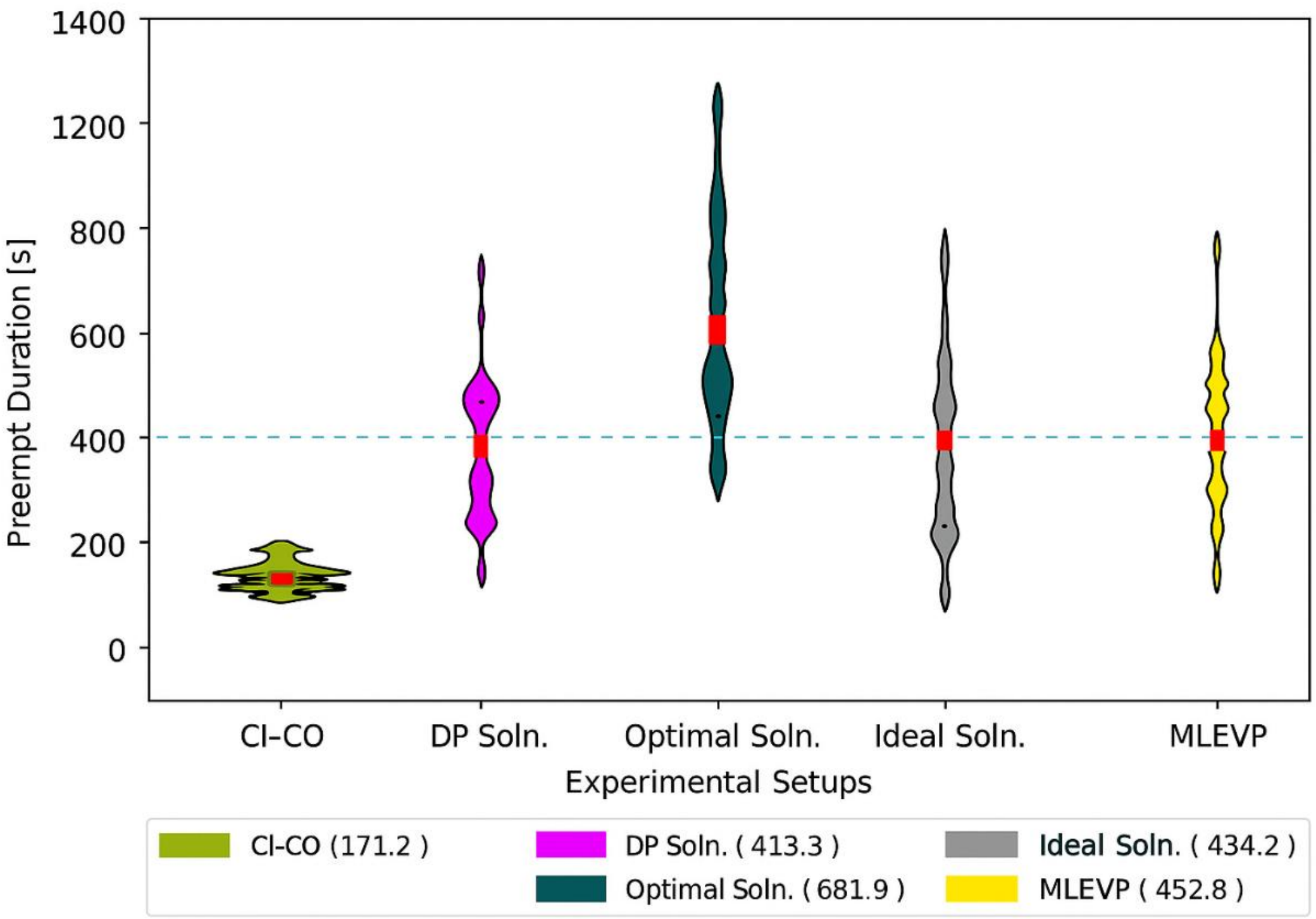

**Fig. 6.** Aggregate preempt duration on seven interchanges at PIB for different experimental strategies.

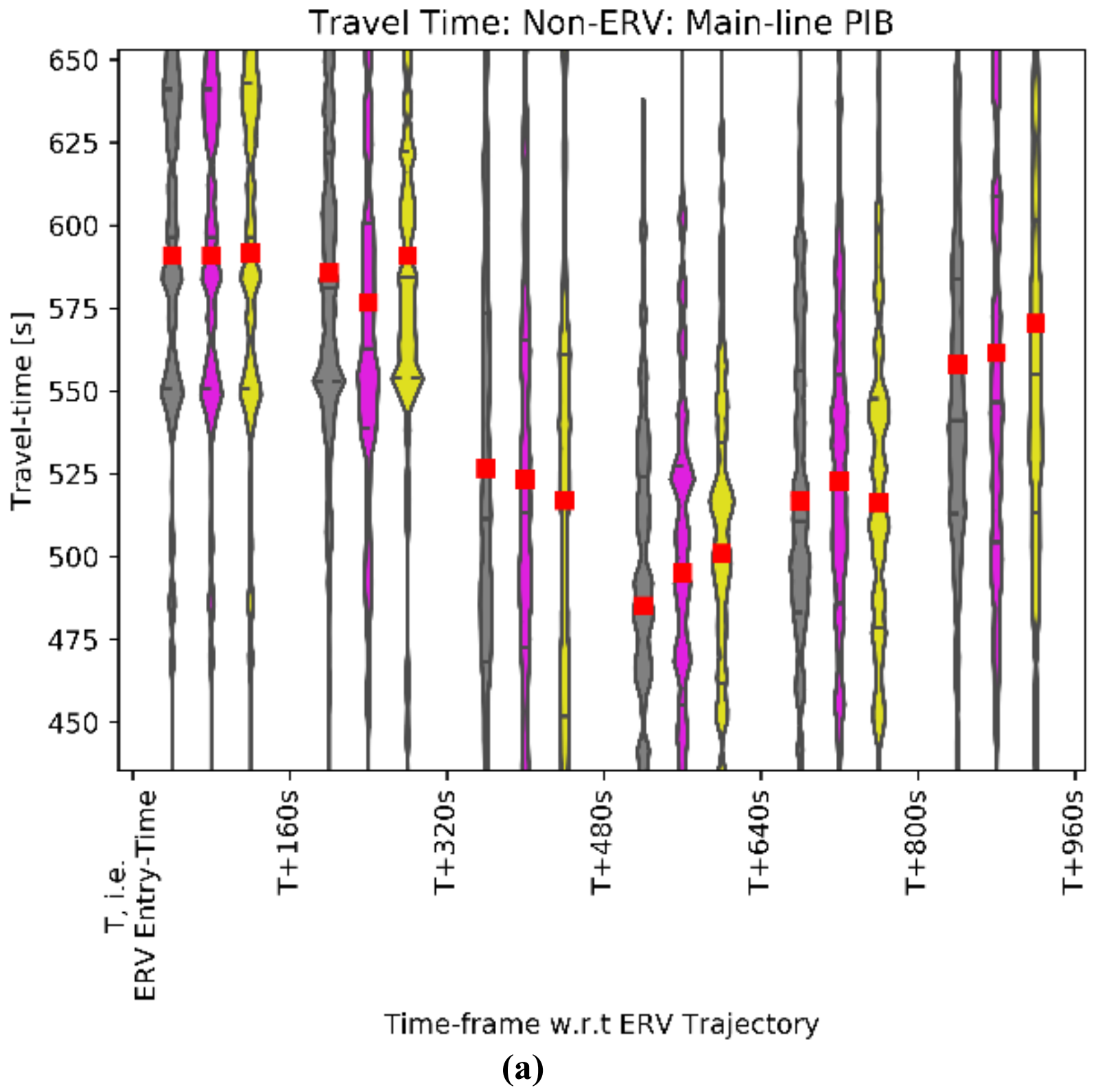

**(a)**

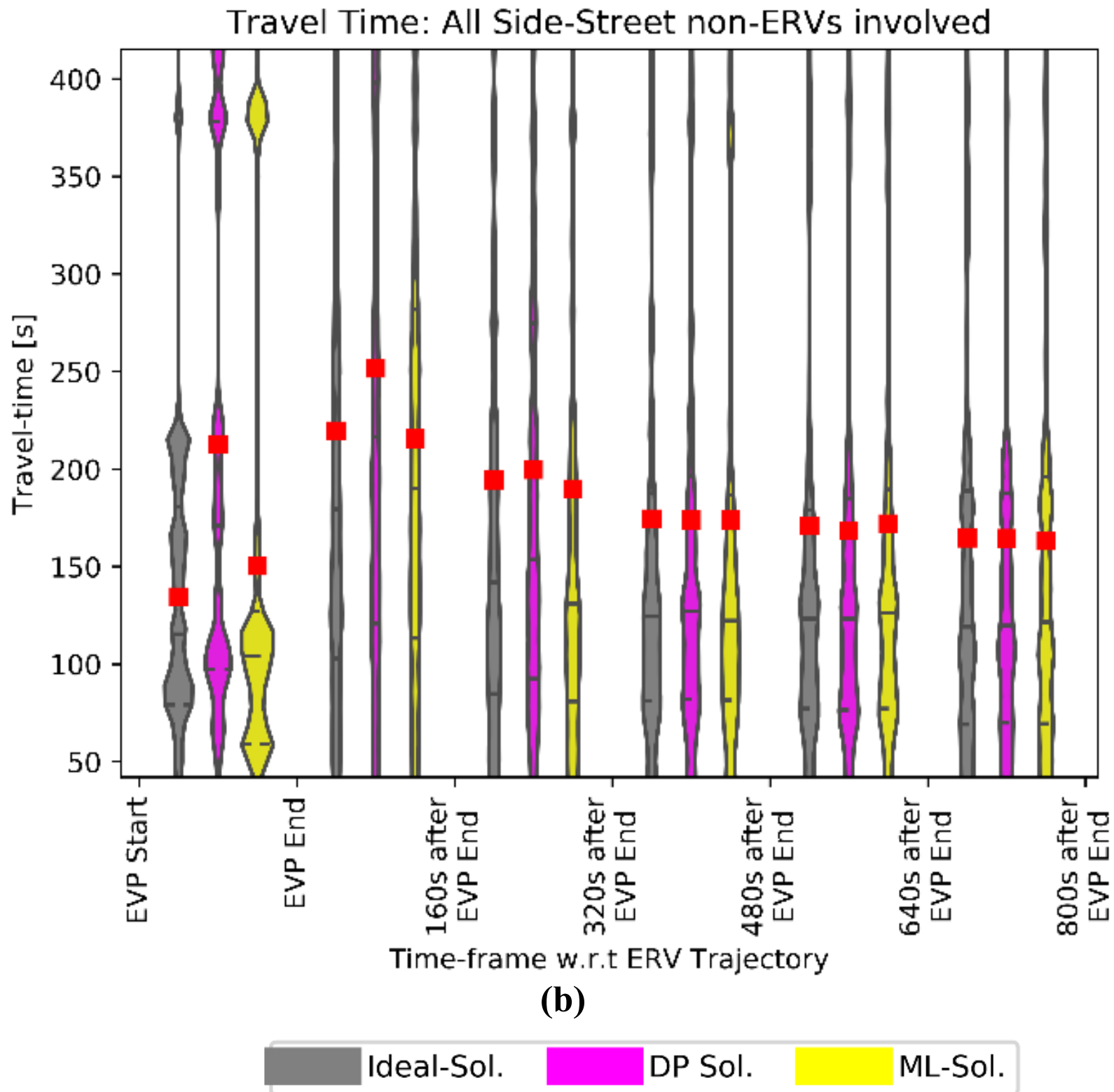

**Fig. 7.** Non-ERV travel time variation with respect to ERV trajectory or preemption event: (a) along main-line PIB and (b) all side-street vehicles since an EVP event.

*B. MLEVP Run Time Discussion*

Although the MLEVP strategy is shown to perform better than other strategies, it is also necessary to test the field applicability of the model, where model execution must be faster than wall-clock time. To estimate the run-time efficiency, experiments are run for both the ring barrier controller (RBC)-driven implementation (i.e., VISSIM's internal controller) and a SIL-simulation (SILS) implementation.

*RBC-Driven Approach*: First, the simulation models are run with and without the MLEVP strategy integration to obtain an estimate of the MLEVP strategy impact on run time. In this experiment, the MLEVP model was run concurrently with the simulation to record the EVP activation times. The simulation conditions were then replicated by triggering EVP at the same time. This approach allowed for an accurate isolation and measurement of any lag introduced by running the MLEVP concurrently with the simulation.

*SILS Approach*: In addition to using the RBC signal controllers in VISSIM, a SILS was tested where each controller is managed using the MaxTime controller emulator, the software used on the controllers in the field. Details on the implementation of the SILS may be found in Roy [41]. The SILS runs use the same scenarios as the RBC approach. Fig. 8 illustrates how both the RBC-driven approach and the SILS approach perform when comparing simulation time to the corresponding wall-clock time, with and without the MLEVP model being active.

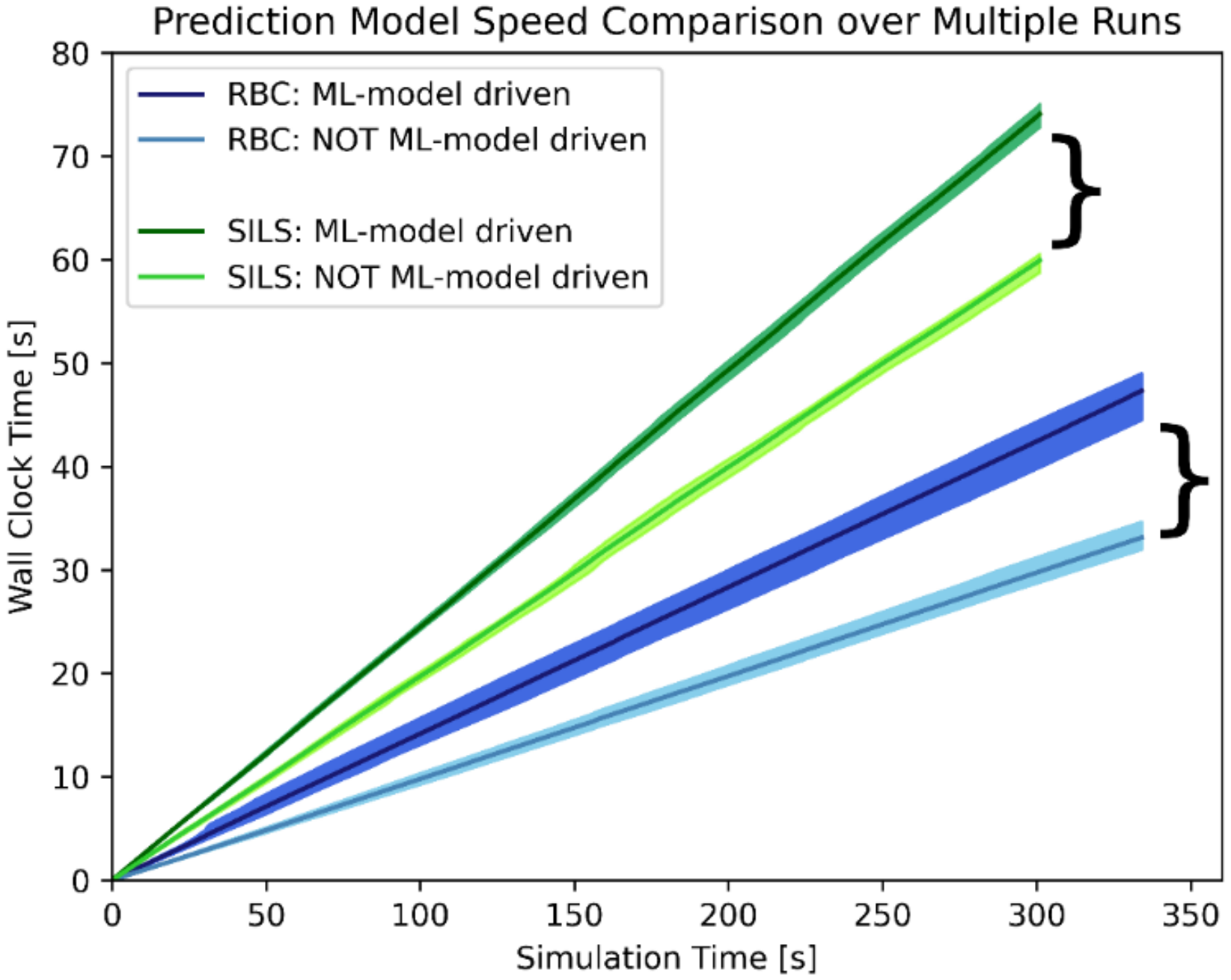

**Fig. 8.** VISSIM® simulation run time in comparison to wall-clock time, with both VISSIM®-RBC and SILS systems.

For comparison in Fig. 8, the same 5 min of simulation time, an hour into the simulation run, are extracted from both the with- and without-MLEVP-model runs. The darkened lines in the plot show the average trend lines for simulation time vs. wall-clock time over 10 replicate runs. The shades around the dark lines show minimum and maximum variation in simulation speed at different instances. All runs are made on a workstation computer with an Intel® Core ™ i7-7700 CPU @ 3.60 GHz and 16 GB of RAM. All trend lines were approximately linear for the entire EVP call period.

For the SILS model runs, the simulation typically executes five times faster than real time (i.e., 1 s of simulation takes 0.2 s of wall-clock time to execute) when the MLEVP strategy is not used to trigger EVP and the EVP is triggered by the SILS signal controller logic. The simulation speed drops slightly to 4× the real-time speed when the MLEVP algorithm is engaged to trigger EVP. In the RBC-driven VISSIM simulation runs, the simulation executes much faster, at 12× real-time speed without the ML model and 8× real-time speed with MLEVP. In actual field deployment, the load on the computer will be significantly lower than the current experiment since the computer will not need to perform the computations related to the simulation but just the computations related to the ML model. However, other delays might manifest in the form of communications delays or delays related to lower performance CPUs used in the field hardware. Nonetheless, this experiment provides a relative measure of the impact of the MLEVP strategy under different computational loads for the different simulation models and provides evidence that the MLEVP strategy can perform faster than real time on retail hardware and can be considered a reasonable candidate for field deployment.

## V. Discussion

This study demonstrates the benefits of using EVP, through the development of an ML model, MLEVP, that uses currently available field data. The MLEVP-based strategy provides a methodology that works with the traditional data while it leverages some facets of CV technology that are less onerous to deploy, such as limited deployment of CV units at intersections and on board the emergency vehicles. This allows for immediate leveraging of CV technology before widespread general deployment. One of the major challenges of using ML models is meeting the significantly heavy data requirements for model training. This effort demonstrated the feasibility of using a simulation platform for the development of a ML model training data for EVP operations. Multiple iterations of simulations were used to generate solutions with varying parameters. MLEVP was trained to learn characteristics from the available set of solutions using factors derived from detector data and signal information. One significant advantage of the proposed approach is that the simulation runs can be easily distributed over multiple compute units to run in parallel and cut down on the longer training time typically involved in other methods such as reinforcement learning. As demonstrated in Fig. 5, the developed MLEVP strategy not only provides significant improvement over non-EVP and traditional EVP approaches but also provides performance similar to known optimal solutions. The MLEVP strategy was also demonstrated to perform faster than real time on retail hardware and can be considered a reasonable candidate for field deployment.

Future efforts will include relaxing the requirement for a minimum distance at which a preempt call is placed and sensitivity tests to explore the robustness of the model and demonstrate the potential for successful application of the model across different scenarios and networks. For longer trips, where multiple ERV routes are possible, considerations for different pathways might also be considered. Lastly, it is assumed that the underlying simulation properly captures the behavior of general traffic when an active emergency vehicle is present. Future efforts will test the robustness of this assumption and potentially include new or calibrated general traffic vehicle-driver models that account for the presence of an active ERV.

## VI. ACKNOWLEDGMENT

The information, data, or work presented here was funded in part by the Gwinnett Department of Transportation and the Center for Transportation Equity Decisions & Dollars (CTEDD), a UTC funded by the USDOT. The authors thank Gwinnett DOT and CTEDD for supporting this research. The authors would like to thank Mr. Tom Sever, Mr. Alex Hoefflich, and Mr. Ken Keena from Gwinnett DOT for their unwavering support throughout the project. The authors also thank the US Department of Energy, Office of Energy Efficiency and Renewable Energy, Vehicle Technologies Office, Energy Efficient Mobility Systems EEMS program, project Real-Twin (EEMS114) for the partial support and funding.